\documentclass[sigconf]{acmart} 
\AtBeginDocument{%
  }

\copyrightyear{2026}
\acmYear{2026}
\setcopyright{cc}
\setcctype{by}
\acmConference[CCS '26]{Proceedings of the 2026 ACM SIGSAC Conference on Computer and Communications Security}{November 15--19, 2026}{The Hague, Netherlands}
\acmBooktitle{Proceedings of the 2026 ACM SIGSAC Conference on Computer and Communications Security (CCS '26), November 15--19, 2026, The Hague, Netherlands}
\acmDOI{10.1145/3830454.3846416}
\acmISBN{979-8-4007-2871-6/2026/11}


\usepackage{multirow}   
\usepackage{enumitem}   
\usepackage{xspace}     
\usepackage{cleveref}   
\usepackage{balance}    

\allowdisplaybreaks

\setlist[itemize]{leftmargin=*,noitemsep, topsep=0pt}
\setlist[enumerate]{leftmargin=*,noitemsep, topsep=0pt}

\newcommand{\ie}{\textit{i.e., }}
\newcommand{\eg}{\textit{e.g., }}

\newcommand{\mypara}[1]{\smallskip\noindent\textbf{#1.} \xspace}
\newcommand{\mysubpara}[1]{\smallskip\noindent\textit{#1.} \xspace}

\newcommand{\fsus}{f_{\mathrm{sus}}}     
\newcommand{\shdist}{f^{\mathrm{dist}}}  
\newcommand{\shind}{f^{\mathrm{ind}}}    

\begin{document}


\title{Poster: A Preliminary Study of LLM Distillation Inference}



\author{Edward Chen}
\email{edward200907@gmail.com}
\affiliation{%
  \institution{Carmel High School}
  \city{Carmel}
  \state{Indiana}
  \country{USA}
}

\author{Yuntao Du}
\authornote{Corresponding author.}
\email{ytdu@purdue.edu}
\affiliation{%
  \institution{Purdue University}
  \city{West Lafayette}
  \state{Indiana}
  \country{USA}
}


\renewcommand{\shortauthors}{Edward Chen and Yuntao Du}

\begin{abstract}
Unauthorized model distillation, in which a model is trained on the outputs of a proprietary large language model (LLM), is a growing threat to model providers.
We study \textit{distillation inference}: determining whether a suspect model was distilled from another model or trained independently.
We formulate this problem as a hypothesis test and estimate the behavior expected under each hypothesis by training \textit{shadow models}: distilled shadow models learn from the teacher's reasoning traces, whereas independent shadow models learn only from reference answers.
The auditor measures how closely each model predicts the teacher's reasoning outputs and then uses the shadow models to convert the suspect's score into a calibrated $p$-value.
In a preliminary study using Qwen2.5-7B as the teacher and Llama-3.2-3B for the suspects, our test achieves a true positive rate of $1.0$ at a significance level of $0.02$.
These results demonstrate the feasibility of using distillation inference to detect distillation attacks.
\end{abstract}

\maketitle

\section{Introduction}
\label{sec:intro}

As large language models (LLMs) grow in capability and commercial value, their providers face a growing threat from \textit{unauthorized distillation attacks}, in which an adversary trains its own model on a proprietary model's outputs to copy its capabilities.
In February 2026, Google~\cite{google2026gtig} and Anthropic~\cite{anthropic2026distillation} both reported a sharp rise in such attacks, which raise legal, intellectual property, and national security concerns~\cite{sharma2025ip}.
These reports highlight the urgent need to determine whether a suspect model was distilled from a specific teacher, a task we call \textit{distillation inference}.

Research on distillation inference has grown rapidly over the past year~\cite{lee2025quantification,wadhwa2025teachers,shi2025detection}.
Existing methods measure the suspect's behavioral similarity to the teacher and report a score or ranking over a candidate pool.
They cannot provide what enforcement requires: a statistical guarantee for a specific teacher--suspect pair, \ie a decision with a $p$-value and a bounded false positive rate (FPR).

To address this, we frame distillation inference as a hypothesis test: is the suspect's behavior closer to a model distilled from the teacher, or to a model trained independently?
Since these two behavioral distributions cannot be observed directly, we approximate them with \textit{shadow models}~\cite{shokri2017membership,carlini2022lira}: \textit{distilled shadow models} fine-tuned on the teacher's chain-of-thought (CoT) and \textit{independent shadow models} fine-tuned on reference answers.
We then score every model with three signals of how well it predicts the teacher's CoT token by token. 
We found that per-instance signals are too weak for reliable distillation auditing; we instead \emph{aggregate} the evidence across the entire audit set into a single per-model likelihood-ratio statistic that yields a calibrated $p$-value and a controlled FPR.

We evaluate our method with Qwen2.5-7B-Instruct as the teacher and Llama-3.2-3B as the base model for the suspects and shadows. At a significance level of $\alpha=0.02$, the test detects every distilled suspect and never flags an independently trained model.

\section{Problem Setup}
\label{sec:setup}

The \textit{teacher} $T$ is a known, typically proprietary model.
The \textit{suspect} $\fsus$ is a published model whose provenance is in question: it was produced either by \textit{distilling} from $T$ or by \textit{independent} training without access to $T$.
The \textit{auditor} distinguishes between these possibilities using an \textit{audit set} $D_{\mathrm{audit}}=\{(x_i,y_i)\}_{i=1}^{N}$ of $N$ questions $x_i$ with answers $y_i$. 
Querying $T$ with $x_i$ yields the teacher's CoT $t_i$.

\mypara{Distillation Inference Protocol}
The audit proceeds as follows:
\begin{enumerate}
  \item The teacher's owner trains $T$ and deploys it with query access.
  \item The suspect's owner curates $D_{\mathrm{train}}$, produces $\fsus$ by distilling from $T$ on this dataset or by training it independently.
  \item The teacher's owner submits $D_{\mathrm{audit}}$ to the auditor.
  \item The auditor runs its algorithm on $D_{\mathrm{audit}}$ using access to $T$ and $\fsus$, outputting a decision $\hat{y}\in\{0,1\}$ ($\hat{y}=1$ means evidence of distillation) with a calibrated $p$-value.
\end{enumerate}

\mypara{Audit Setting}
We consider a setting in which the teacher's owner holds logs of queries submitted by the suspect's owner and uses them as the audit set, \ie $D_{\mathrm{audit}}\subseteq D_{\mathrm{train}}$. 
This arises when the API logs containing the distillation queries can be identified~\cite{google2026gtig,anthropic2026distillation}.

\mypara{Access Levels}
The auditor has query access to the teacher to collect $t_i$.
For the suspect, we assume either \textit{grey-box} access (\ie next-token probabilities, available for any open-weight release or log-probability API) or \textit{black-box} access (\ie generated text only).

\section{Methods}
\label{sec:methods}

\mypara{Overview}
Our method instantiates the audit as a binary hypothesis test: is $\fsus$'s behavior on $D_{\mathrm{audit}}$ more consistent with a model distilled from $T$ (the alternative hypothesis $H_1$) or with one trained independently (the null hypothesis $H_0$)?
Since the two distributions are unknown, we estimate them in three stages: fine-tune distilled shadows and independent shadows models (Stage~1); score every model with three signals of how closely it tracks the teacher's CoT (Stage~2); aggregate the per-instance signal into a likelihood-ratio test calibrated by the shadows, returning a $p$-value and a verdict at a controlled FPR (Stage~3).

\mypara{Stage 1: Training Shadow Models}
The auditor fine-tunes $M$ shadow models from each class~\cite{carlini2022lira}. These models share the same base model, audit set, and training procedure and differ \textit{only} in their supervision targets.
A distilled shadow $\shdist_m$ is fine-tuned with Low-Rank Adaptation (LoRA)~\cite{hu2022lora} to reproduce the teacher's CoT $t_i$ for each $x_i$. 
In contrast, an independent shadow $\shind_m$ is fine-tuned on the reference answer $y_i$ without access to $T$, representing a model trained without distillation.
Each model is trained using the randomly sampled half of the $D_\text{audit}$. 
Thus, every pair of distilled and independent shadows differs only in its training target (\ie $t_i$ or $y_i$).
This design controls for confounding factors because knowledge shared by both classes cancels out in shadow models.


\mypara{Stage 2: Extracting Distillation Signals}
Every shadow model $f$ is scored in a \textit{single teacher forward pass} over the teacher's CoT: for instance $i$ with CoT tokens $t_{i,1{:}L_i}$, the pass yields the next-token distribution $p^{f}_{i,\ell}=f(\cdot\mid x_i,\,t_{i,<\ell})$ at every position $\ell$.
Averaging over the $L_i$ positions, we extract three signals presenting how closely $f$ tracks the teacher.
\begin{itemize}
  \item \textbf{Log-probability (grey-box).}
  It measures the mean log-likelihood of the teacher's tokens under $f$:
  $g_{\mathrm{lp}}(f,i)=\tfrac{1}{L_i}\textstyle\sum_\ell \log p^{f}_{i,\ell}(t_{i,\ell})$. A distilled model was trained to reproduce this text, so it scores the text higher than an independent model that learned a different but equally valid solution.
  \item \textbf{Predictive entropy (grey-box).}
  We compute the mean entropy as $g_{\mathrm{ent}}(f,i)=\tfrac{1}{L_i}\textstyle\sum_\ell \mathbb{H}(p^{f}_{i,\ell})$ to measure the certainty of model about the outputs.
  \item \textbf{Token agreement (black-box).}
  We compute the fraction of positions at which $f$'s greedy token $\hat{t}^{f}_{i,\ell}=\arg\max_v p^{f}_{i,\ell}(v)$ matches the teacher's token as $g_{\mathrm{ta}}(f,i)=\tfrac{1}{L_i}\textstyle\sum_\ell \mathbf{1}[\hat{t}^{f}_{i,\ell}=t_{i,\ell}]$. This signal directly measures whether $f$ would emit the teacher's tokens when decoding on its own. 
\end{itemize}

\mypara{Stage 3: Performing Distillation Inference}
Our test adapts the per-example construction of the Likelihood Ratio Attack (LiRA)~\cite{carlini2022lira}. It has two levels: a \textit{per-instance} score and a \textit{model-level} statistic, both calibrated using the shadows.

\mysubpara{Per-instance distributions and the calibrated score}
For instance $i$, the values from the distilled shadows, $\{g(\shdist_m,i)\}_{m}$, and the independent shadows, $\{g(\shind_m,i)\}_{m}$, characterize how the signal behaves under the two hypotheses. We summarize the two distributions with Gaussians $\mathcal{N}(\mu_{\mathrm{dist},i},\sigma^{2}_{\mathrm{dist},i})$ and $\mathcal{N}(\mu_{\mathrm{ind},i},\sigma^{2}_{\mathrm{ind},i})$, whose parameters are the means and unbiased variances over the $M$ shadows.
The evidence provided by model $f$ for instance $i$ is the log-likelihood ratio (LLR) of its value under the two distributions:
\begin{equation}
\lambda_i(f)=\log
\frac{p\big(g(f,i)\mid\mathcal{N}(\mu_{\mathrm{dist},i},\,\sigma^{2}_{\mathrm{dist},i})\big)}
     {p\big(g(f,i)\mid\mathcal{N}(\mu_{\mathrm{ind},i},\,\sigma^{2}_{\mathrm{ind},i})\big)}.
\label{eq:llr}
\end{equation}
The \textit{calibration} matters because $g(f,i)$ alone conflates specificity to the teacher with per-instance difficulty (every model scores high on an easy instance), whereas $\lambda_i$ measures $f$ \textit{relative to} both classes on that instance.

\mysubpara{Model-level statistic and its null/alternative}
Summing over the audit set gives the statistic $\Lambda(f)=\sum_{i=1}^{N}\lambda_i(f)$, the most powerful test at any fixed FPR under this Gaussian model~\cite{neyman1933problem}.
Its distribution under each hypothesis is estimated from the shadows using a leave-one-out procedure. Each shadow is scored against Gaussians fitted to the \textit{other} shadows in its class, yielding one draw from $H_1$ (distilled) or $H_0$ (independent). Thus, no model is evaluated against a distribution that it helped define.
The suspect, which is trained separately from the shadows, is scored against Gaussians fitted to all shadows.

\mysubpara{Audit Decision}
The $p$-value ranks $\Lambda(\fsus)$ against the scores of the independent shadows,
\begin{equation}
p=\frac{1+\#\{\,m:\Lambda(\shind_m)\ge\Lambda(\fsus)\,\}}{M+1},
\label{eq:pvalue}
\end{equation}
\ie the estimated frequency with which an independent model matches the teacher at least as closely as the suspect does. The decision is $\hat{y}(\fsus)=\mathbf{1}[p<\alpha]$.
We estimate the FPR by testing each independent shadow against a threshold fitted without that shadow, so the estimate changes in increments of $1/M$. 
Both quantities become more precise as $M$ increases.

\section{Preliminary Evaluation}
\label{sec:results}

\begin{figure}[!tb]
  \centering
  \includegraphics[width=\columnwidth]{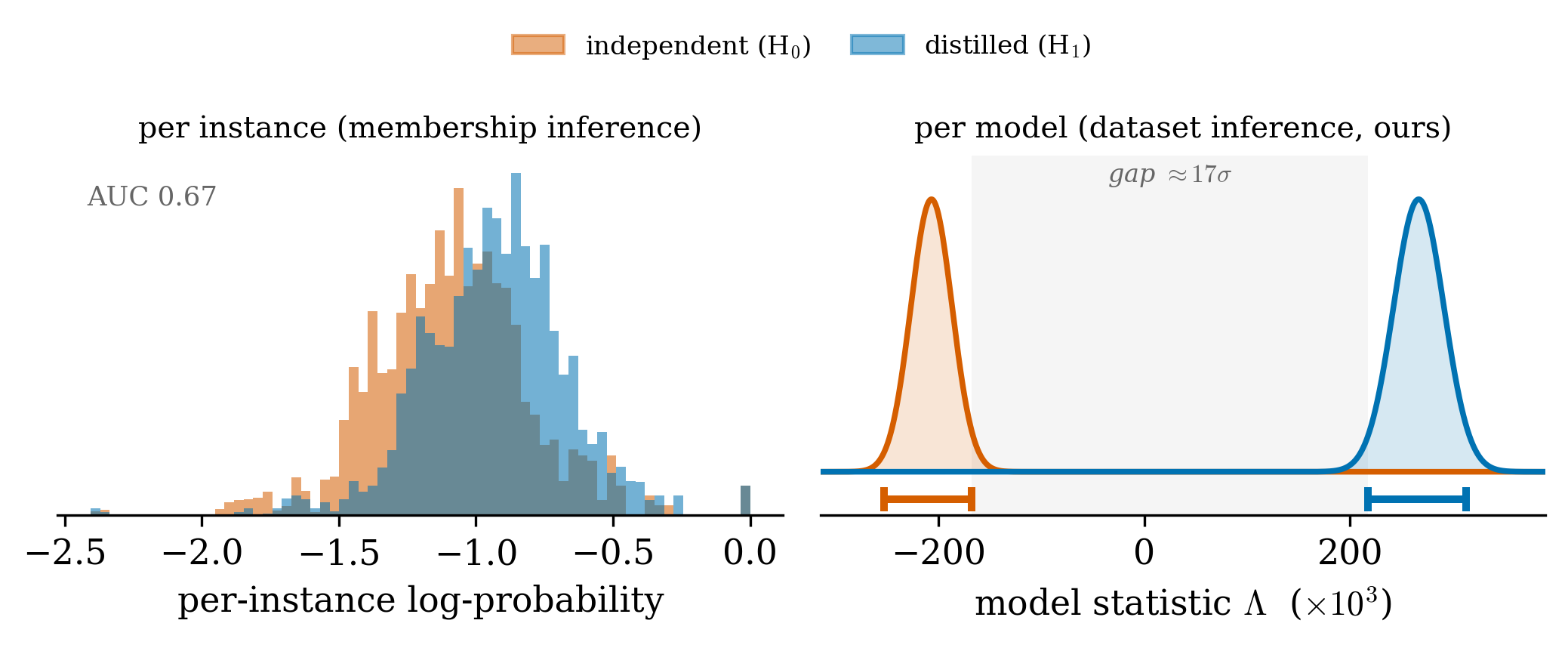}
  \caption{Per instance (left), the log-probability signal overlaps across the two classes, so
  membership inference cannot separate them (AUC $0.67$). Aggregated to one statistic $\Lambda$ per
  model (right), the distilled ($H_1$) and independent ($H_0$) populations are fully separated,
  so our methods can faithfully detect distillation.}
  \Description{Two panels. Left: the per-instance log-probability histograms of independent and distilled shadow models overlap heavily (AUC 0.67). Right: the per-model statistic Lambda forms two well-separated bell curves for the independent and distilled populations, with a gap of about 17 standard deviations.}
  \label{fig:signals}
\end{figure}

\mypara{Experimental Setup}
The teacher $T$ is Qwen2.5-7B-Instruct, and the base model for the suspect and all shadows is Llama-3.2-3B. The audit set $D_{\mathrm{audit}}$ contains $N=500$ instances from CoT-Collection~\cite{kim2023cotcollection}, and we train $M=50$ shadows per class.
All models are fine-tuned with LoRA (rank $8$, scaling factor $16$, and dropout $0.05$ on the attention projections) for three epochs. We retain the checkpoint with the lowest loss on the held-out half and build all prompts with a shared formatter to rule out formatting confounds.
Suspects are trained separately from the shadows using the same procedure: distilled suspects learn from the teacher's CoT, whereas independent suspects learn from reference answers.
We train these models to achieve similar utility as their base model by evaluating them on a separate held-out dataset.

\mypara{Baselines and Evaluation}
We contrast two ways of turning the per-instance signals of \Cref{sec:methods} into a verdict.
The \textit{membership-inference} baseline~\cite{shokri2017membership} treats each instance in isolation: it thresholds a single instance's signal to decide whether that model is distilled.
We report audit performance with AUC, balanced accuracy, and TPR at $0\%$ and $5\%$ FPR. 


\mypara{Results}
\emph{(i) Membership inference does not detect distillation.} On individual instances the signals barely separate the classes (AUC $0.64$--$0.67$), and they are useless in the low-FPR regime that enforcement requires: TPR at $0\%$ FPR is essentially zero ($\le 0.003$) and stays below $0.10$ even at $5\%$ FPR (\Cref{tab:signals}, ``per-instance''; \Cref{fig:signals}, left).
\emph{(ii) Our method separates the model populations perfectly.} Aggregating the same signals across $D_{\mathrm{audit}}$ lifts every metric to $1.000$ (\Cref{tab:signals}, ``per-model''; \Cref{fig:signals}, right): the $M=50$ distilled and independent populations do not overlap, so the audit flags every distilled suspect (TPR $1.0$) and clears every independent one. 
\emph{(iii) Shadow calibration certifies the decision.} 
A raw aggregate score separates the populations but cannot bound the false-positive rate. 
Calibrating $\Lambda$ against the shadow populations turns it into a hypothesis test with a certified operating point: at $M=50$ the $p$-value reaches its floor $1/(M{+}1)=0.020$ and the leave-one-out FPR is $\approx 1/M = 0.020$, both shrinking monotonically with $M$ while detection stays perfect (\Cref{tab:scaling}). 

\begin{table}[t]
  \centering
  \setlength{\tabcolsep}{4pt}
  \caption{Membership inference on individual instances barely separates distilled from independent models, whereas aggregating the same signal across the audit set (dataset inference, our method) separates the model populations perfectly. bal-acc = balanced accuracy.}
  \label{tab:signals}
  \resizebox{\columnwidth}{!}{%
  \begin{tabular}{llrrrr}
    \toprule
    Signal & & AUC & bal-acc & TPR@5\%FPR & TPR@0\%FPR \\
    \midrule
    \multirow{2}{*}{log-prob}
      & per-instance (MIA) & 0.672 & 0.518 & 0.086 & 0.000 \\
      & per-model (ours)   & \textbf{1.000} & \textbf{1.000} & \textbf{1.000} & \textbf{1.000} \\
    \midrule
    \multirow{2}{*}{entropy}
      & per-instance (MIA) & 0.672 & 0.524 & 0.098 & 0.000 \\
      & per-model (ours)   & \textbf{1.000} & \textbf{1.000} & \textbf{1.000} & \textbf{1.000} \\
    \midrule
    \multirow{2}{*}{token agreement}
      & per-instance (MIA) & 0.635 & 0.512 & 0.073 & 0.003 \\
      & per-model (ours)   & \textbf{1.000} & \textbf{1.000} & \textbf{1.000} & \textbf{1.000} \\
    \bottomrule
  \end{tabular}%
  }
\end{table}

\mypara{Impact of Model Architecture}
The main experiment uses a Qwen teacher and a Llama suspect. 
We find that a same-family pair (Qwen$\to$Qwen), which shares a tokenizer and output format, produces \textit{larger} separation between the two distributions. This result rules out model-formatting differences between families (\eg whitespace patterns in generation) as confounding factors.

\mypara{Mitigation}
We also evaluate an evasion strategy in which the suspect's owner trains the suspect on a \textit{paraphrased} teacher CoT. We find the inference statistic decreases from original CoT to teacher-generated paraphrase, and decreases again for Mistral-generated paraphrase.
All suspects remain flagged at $\alpha=0.05$ (these ablation pools use $M=20$, where the $p$-floor is $1/(M{+}1)\approx0.05$), although the paraphrase-trained suspects enter the gap between the populations. Replacing the distilled shadows with a matching population trained on paraphrased text restores a clean audit, as expected.

\begin{table}[t]
  \centering
  \setlength{\tabcolsep}{8pt}
  \caption{More shadow models tighten the certified operating point: the $p$-value floor $1/(M{+}1)$ and the FPR ($\approx 1/M$) both shrink with $M$, while per-model detection stays perfect (AUC $=1.000$ throughout).}
  \label{tab:scaling}
  \begin{tabular}{rcc}
    \toprule
    $M$ & $p$-floor & FPR \\
    \midrule
    5  & 0.167 & 0.200 \\
    10 & 0.091 & 0.100 \\
    20 & 0.048 & 0.050 \\
    40 & 0.024 & 0.025 \\
    50 & 0.020 & 0.020 \\
    \bottomrule
  \end{tabular}
\end{table}

\section{Conclusion and Limitations}
\label{sec:conclusion}


In this paper, we frame distillation inference as a hypothesis test and propose a likelihood-ratio test calibrated with shadow models.
In a preliminary study, our method detects every distilled suspect model without any false positives, providing a statistical bound on FPR for enforcement.
Our study has three limitations.
First, it requires fine-tuning a pool of shadow models, which may be expensive in practice given the size of frontier LLMs.
Second, it assumes that the audit set is a subset of the suspect's training data, which may be hard to establish in practice.
Third, it assumes that the suspect is trained with the same procedure as the shadow models.
Relaxing the last two assumptions is an important direction for future work.


\bibliographystyle{ACM-Reference-Format}
\balance   
\bibliography{ref}

\end{document}